\documentclass[twocolumn,showpacs,preprintnumbers,amsmath,amssymb]{revtex4-1}

\usepackage{graphicx}
\usepackage{dcolumn}
\usepackage{bm}

\begin{document}

\preprint{APS/123-QED}

\title{Nonequilibrium evolution thermodynamics theory}

\author{Leonid S. Metlov}

 \email{lsmet@fti.dn.ua}
\affiliation{Donetsk Institute of Physics and Engineering, Ukrainian
Academy of Sciences,
\\83114, R.Luxemburg str. 72, Donetsk, Ukraine
}%

\date{\today}

\begin{abstract}
Alternative approach for description of the non-equilibrium phenomena arising in solids at a severe external 
loading is analyzed. The approach is based on the new form of kinetic equations in terms of the internal and 
modified free energy. It is illustrated by a model example of a solid with vacancies, for which there is a 
complete statistical ground. The approach is applied to the description of important practical problem – 
the formation of fine-grained structure of metals during their treatment by methods of severe plastic deformation.   
In the framework of two-level two-mode effective internal energy potential model the strengthening curves unified 
for the whole of deformation range and containing the Hall-Petch and linear strengthening sections are calculated. 
\end{abstract}

\maketitle

\section{Introduction}

An idea to use an additional variable for account of internal microstructure changes in polyatomic gases was proposed 
by Herzfeld and Rice in 1928 \cite{hr28}. This idea has got development in other numerous researches of that 
time \cite{r33,k33,lt36,ml37}. In 1937 Landau applied it to describe phase transitions in solids \cite{l37}. In the post-war 
years Landau in cooperation with Khalatnikov and Ginzburg for description of phase transitions offered some kinetic 
equations, describing the evolution of order parameter \cite{lk54,gl50}. In modern science this direction is presented 
by the theory of the phase fields \cite{akv00,kkl01,esrc02,lpl02,rbkd04,akegny06,s06,rjm09}. Unlike the theory of Landau 
the parameter of order in the theory of the phase fields is not so strict, however, this does not hinder in getting results 
which rather well coincide with the behavior of the real systems.

To complete the picture, note that another direction in the description of such phenomena was offered in 1967 
by Coleman and Gurtin, who complemented the primary idea by the elements of rational mechanics \cite{cg67}.
This direction was continued in Ref. \cite{d70,pb69,m97,v01,v02,kt05,k05}. In Ref. \cite{m97} an additional variable 
(«order parameter»), meaning the scalar density of embryonic microcracks, was introduced to describe the initial stage 
of destruction of a quasi-brittle solid. Later in the work by Peter Van \cite{v01} the application of internal state 
variables approach to the problem of destruction of cracking materials was examined. For the description of crackness 
level an additional vector variable $\vec{\alpha}$ is introduced, which is related to thermodynamic conjugated 
variable $\vec{A}$.

Between both approaches there is no an impenetrable border, they use one and the same fundamental idea. 
The difference is that the internal state variables approach one comes from the principles and postulates of mechanics 
of continuous media, while the approach by Landau is based on more general energetic principles, which, however, need 
further development. Interest which is sometimes shown in comparison of these two approaches is therefore clear \cite{v02}.

A further approach of mesoscopic nonequilibrium thermodynamics is currently developed for investigation of soft-matter 
problem with mesoscopic structural elements \cite{rrp98,rg03,rvr05,pmrl11}. The method addresses the system’s entropy 
production, which is a detectable quantity at the mesoscopic level of condensed-matter organization; it also reveals 
effectively the kinetic-thermodynamics properties and peculiarities of the corresponding dynamic matter arrangements.

The present article is devoted to the development of the Landau approach, which in the present version appears under 
the name of non-equilibrium evolutional thermodynamics (NEET). In part II, the basic postulates of NEET are given, 
which are grounded by the results attained currently. In part III the approach is applied to the description of 
defect kinetics during metal treatment by severe plastic deformation (SPD). In part IV summarizing conclusions are given.

\section{Postulates of NEET}

The first important difference of NEET from the classic scheme treating the problem is the introduction of the set of 
non-equilibrium thermodynamics potentials \cite{m07,m07c}. Up to the present, basic, if not unique, stress was laid on 
the use of the free energy. The free energy is still a major function, which in energetic expression and logarithmic scale 
is a value inverse to the probability distribution function (PDF) of the system states \cite{s89}. A postulate is 
well known in statistical physics that the most probable states near a maximum of PDF, that is, automatically, near a 
minimum of the free energy \cite{ll69} are realized only.

In this context the use of the free energy is very comfortable, because the most probability (stationary) state can be 
found from the universal principle of a minimum of the free energy. At the same time, in structural physics not for all 
of the physical systems the PDF is certain, and, consequently, the free energy on its basis. Mainly, one uses a 
phenomenological generalization of the free energy, when the application of principle of minimum is not grounded 
statistically. Applicability of this principle is well proven most strictly for solids with vacancies, at the same time, 
for solids with other types of defects (dislocations, grain boundaries etc) PDF is unknown, and, consequently, 
applicability of this principle is not grounded strictly. Therefore a solid with vacancies can be a good model for 
approbation of ideas, having general character \cite{m10,m10p,m11}.

If the equilibrium (stationary) state of the system is known, in the case of small deviation of the system from that states, 
the evolution (relaxation) of the system can be written down as a condition that the system tends to the equilibrium state 
with a speed the greater the stronger the deviation. In terms of the free energy this condition results in the well known 
Landau-Khalatnikov evolution equation \cite{lk54} or in the case of the distributed systems in the Ginzburg-Landau 
equation \cite{gl50}. At the same time, there is no obstacle to write down the evolution of the system in terms of 
other thermodynamic potentials, for example, in terms of the internal energy.

The internal energy is the clearest physically determined energy of the system; it is basic in both thermodynamics and 
physics in general. This energy is included in formulation of the first law of thermodynamics; it is universal for both 
the equilibrium and non-equilibrium states. In addition, the generalized thermodynamic force is determined through it not 
phenomenologically, as, for example, for the free energy, but fully strictly within the framework of statistical 
consideration. Indeed, PDF for a solid with vacancies looks like \cite{s89,g97}.
  \begin{equation}\label{b1}
f(n)=CW\exp(-\dfrac{U(n)}{k_{B}T})=C\dfrac{(N+n)!}{N!n!}\exp(-\dfrac{U(n)}{k_{B}T}),
  \end{equation}
where $C$ is a normalizing constant, $W$ is the thermodynamic probability, $U(n)$ is the internal energy, 
$N$ is the number of atoms in a solid, $n$ is the number of vacancies, $k_{B}$ is the Boltzmann's constant, 
$T$ is a temperature. A pre-exponential multiplier describes combinational, that is entropic, part of the distribution 
function, related to degeneration of macrostates. The exponent describes a restrictive part of the distribution function, 
related to overcoming the potential barriers between microstates. The most probable state is determined by a condition
$\partial f(n)/\partial n$, from where, in accordance with (1), most naturally appears a variable
  \begin{equation}\label{b2}
u=\frac{\partial U}{\partial n}.
  \end{equation}
It has a sense of the average energy of a defect (here a vacancy), or of the chemical potential of defects, on the 
other side, Eq. (\ref{b2}) is a typical determination of the generalized thermodynamic force. Here it is not, however, 
postulated, but logically follows from Eq. (\ref{b1}), as a part of the process of determination of the equilibrium state, 
and, consequently, it must enter relaxation equation, as the equation describing system tendency to the equilibrium state. 
Now the evolutional equation in terms of the internal energy can be written down as
  \begin{equation}\label{b3}
\dfrac{\partial n}{\partial t}=\gamma_{n}(\dfrac{\partial U}{\partial n}-u_{e}),
  \end{equation}
where $\gamma_{n}$ is a kinetic coefficient, $u_{e}$ is a value of the vacancy energy in the equilibrium state.

To define the equilibrium value of the vacancy energy and the equilibrium density of vacancies, we all the same must 
address the condition of a maximum of PDF or a minimum of the free energy, and, it would seem, it is simpler to describe 
relaxation in terms of the free energy, immediately using the Landau-Khalatnikov equation. However, Eq. (\ref{b3}) is also 
true, and can also be used. Especially it can be valuable at generalizations to other types of defects, for which 
phenomenological generalization of the free energy is problematic, and at times speculative, whereas the internal energy 
is of the universal nature.

The substantial difference of treating the problem in terms of the internal energy is that the generalized force (\ref{b2}) 
is not equal to zero in the equilibrium state, and, consequently, for the internal energy the extreme principle does not 
work. This «pitfall» can be compensated as the generalized force (\ref{b2}) is determined uniformly for both the equilibrium 
and non-equilibrium states, if the equilibrium energy of the defect is known for us from some sources for the defined values 
of external control parameters, then it is possible to find its value for other parameters easily. This advantage is for 
the first time exposed in the present article.

The second postulate, which distinguishes this approach from traditional one, is the use of the density of defects 
as an independent thermodynamic variable instead of the configuration entropy. Note that this variable is used not 
in parallel with the configuration entropy, but instead of it. Mutually identical dependence between the density of 
defects and the configuration entropy can serve as a foundation for this purpose. In the case of a solid with vacancies 
this one-to-one dependence follows from the fundamental Boltzmann relationship $S_{c} = k_{B}\ln W$ and from definition of 
$W$ in accordance with Eq. (\ref{b1}). For other types of defects this relation is unknown, but it still must be 
mutually identical. It allows to generalize the 1st law of thermodynamics in the form \cite{m10}
  \begin{equation}\label{b4}
dU=V \sigma_{ij}d\varepsilon_{ij}+TdS+\tilde{T}d\tilde{S}+\sum_{l=1}^{N_{def}}\varphi_{l}dh_{l},
  \end{equation}
where $V$ is the volume of the system, $\sigma_{ij}$, $\varepsilon_{ij}$ are stress and elastic deformations tensors, 
$S$ is thermal or, for a solid, oscillation (not configuration) entropy, $\tilde{T}$, $\tilde{S}$ are non-equilibrium 
temperature and entropy, which characterize the dynamic transitional phenomena during the generation and motion of the 
structural defects \cite{m10}, $\varphi_{l}$, $H_{l}$ are energy and density of $l$-kind defects, $N_{def}$ is the 
number of types of the defects. Thus the internal energy is a function of such independent variables as 
$\varepsilon_{ij}$, $S$, $\tilde{S}$ and $H_{l}$, that is, $U=U(\varepsilon_{ij},S,\tilde{S},H_{l})$.

The first two terms in Eq. (\ref{b1}) are changes in the internal energy due to contribution of the elastic stress 
field and equilibrium thermo-motion, the third term characterizes a part of the internal energy, arising due to 
non-equilibrium transient processes (the necessity of its account is grounded in Ref. \cite{m10}), and the last term 
presents a part of the internal energy concentrated in defect subsystems.

Relation (\ref{b2}) in the case of arbitrary number of defect types can be generalized as
  \begin{equation}\label{b5}
\varphi_{l}=\frac{\partial U}{\partial H_{l}},
  \end{equation}
and evolution Eq. (\ref{b3}) accordingly
  \begin{equation}\label{b6}
\dfrac{\partial H_{l}}{\partial t}=\gamma_{l}(\dfrac{\partial U}{\partial H_{l}}-\varphi_{le}).
  \end{equation}

Carrying out transformation of the Legendre type, but with respect to the pair of thermodynamic variables of 
$\varphi_{l}$ and $H_{l}$, we pass to a new thermodynamics potential
  \begin{equation}\label{b7}
\tilde{F}_{l}=U-\varphi_{l}H_{l},
  \end{equation}

It is not hard to show that for this function a relation 
  \begin{equation}\label{b8}
H_{l}=-\dfrac{\partial U}{\partial \varphi_{l}},
  \end{equation}
is just, that is, the new function $\tilde{F}_{l}$ is related to the internal energies of $U$ in the same way as the 
classic (thermal or oscillation) free energy \cite{b64}, if the entropy is formally taken instead the number of 
defects $H_{l}$, and the temperature instead of the energy of defect $\varphi_{l}$. But with the second postulate 
of NEET such the accordance is recognized, that is, the function $\tilde{F}_{l}$ can be interpreted, as a modified 
free energy, when the number of defects of $H_{l}$ is taken as an independent thermodynamic variable. 

If transformation (\ref{b7}) is done for all of types of defects, then such modified free energy will be the function 
of independent variables $\varepsilon_{ij}$, $S$, $\tilde{S}$ and $\varphi_{l}$, that is, 
$\tilde{F}_{l}=\tilde{F}_{l}(\varepsilon_{ij},S,\tilde{S},\varphi_{l})$. Thus, an «own» argument for the internal energy is 
the defect densities, and for the modified free energy – the energy of defects, similar to the case of classic 
thermodynamics, where an «own» argument for internal energy is the entropy, and for free energy is the temperature.

Classical configurational free energy $F_{c} =U-TS_{c}$ is always treated, as one-to-one function of the density of 
defects, while in obedience to the method of definition (Legendre transformation) it must be the function of temperature. 
It follows that it is not a thermodynamics potential in a strict sense, and it is only useful as energy reflection of PDF. 
While the internal energy and the modified free energy, satisfying Eqs. (\ref{b5}), (\ref{b6}) and (\ref{b8}), 
are real thermodynamics potentials, though dissatisfy extreme principle.

\section{Two-level two-mode model of SPD}

Now setting the dependence of the internal energy on its arguments, we fully determine our problem in a thermodynamic 
sense. Let us apply the above approach for solution of a special problem. Producing a fine-grained structure of metals 
by severe plastic deformation (SPD) is presently urgent. At the initial stage of SPD there goes intensive generation 
of dislocations, then next the dislocations serve as a building material for the growth of grain boundaries, that 
results in a finer grain structure. Thus, in the processes of SPD these two types of defects take the main part and 
predetermine the two-level character of the problem \cite{m09}.

\subsection{Evolution equations}

Let us consider a homogeneous problem, setting the internal energy as a polynomial dependence
  \begin{equation}\label{b9}
u=u_{0}+\sum_{l=g,D}^{}(\varphi_{0l}h_{l}-\varphi_{1l}h_{l}^2+\varphi_{2l}h_{l}^3-\varphi_{3l}h_{l}^4)+\varphi_{gD}h_{g}h_{D},
  \end{equation}
where $u_{0}$, $\varphi_{kl}$, $\varphi_{gD}$ are some coefficients, depending on the equilibrium variables of $s$ 
and $\varepsilon_{ij}$, as control parameters
\begin{eqnarray}\label{b10}
\nonumber
u_{0}=\dfrac{1}{2}\lambda(\varepsilon_{ii}^e)^2+\mu(\varepsilon_{ij}^e)^2+\beta s^2,  \\
\varphi_{0l}=\varphi_{0l}^*+g_{l}\varepsilon_{ii}^e
+\dfrac{1}{2}\bar{\lambda}(\varepsilon_{ii}^e)^2+\bar{\mu}(\varepsilon_{ij}^e)^2
-\beta_{l}s,     \\
\nonumber
\varphi_{1l}=\varphi_{1l}^*-2e\varepsilon_{ii}^e.
\end{eqnarray}
For the sake of convenience we passed from the numbers of defects of $H_{l}$ to the densities of corresponding 
variables of $h_{l}$, and similarly $S \rightarrow s$, $U \rightarrow u$.

The fourth-degree polynomial in parentheses can have at the positive values of coefficients $\varphi_{km}$ two maxima 
(two modes). The mode which corresponds to the lower value of defectiveness, in the case of dislocations $l = D$, 
can describe the accidental (homogeneous) distribution of dislocations. The mode, which corresponds to the higher 
value of defectiveness, describes dislocations belonging to the cell structure in this case. We examine only the 
simplified case of the homogeneous distribution of dislocations, that is, $\varphi_{3D}=0$ and $\varphi_{4D}=0$. 

The coefficient of $\varphi_{0g}$ can be considered as a general surface density of the energy of regular (infinity) GB. 
From data of S.A. Firstov, for cold-roll treatment this energy can equal the doubled energy of a free boundary for 
same material. For example, it can make approximately $2\times2 J/m^2$ for copper. The first term   in this context 
is the own energy of the boundary without a contribution from other factors. It is considered, as a well 
full-relaxated grain boundary, that is, as a minimum possible surface energy of GB. According to A.S. Firstov, 
this energy makes $0.15 \div 0.20$ of the energy of free-surface of the same material. That is, for copper this 
energy approximately equals $0.2\times2=0.4 J/m^2$.

Contribution of the second term $g_{g}\varepsilon_{ij}^e$ to the field of compressing hydrostatic stress results 
in the decreasing of GB energy. It is of great physical importance meaning that the grain boundaries are sites of 
density lack distributed along some surface. Exactly these sites give the highest contribution to the energy of 
boundaries. When, due to external pressure, the volume of undensitied sites diminishes, the energy of boundaries 
diminishes, as well as the potential barriers between the stable neighbor states, that results in growing mobility 
of the grain boundaries. If elastic deformation is $0.002$ that for copper corresponds to the level of 
tension $\sim 180 MPa$, the constant gg is to be taken within the limits of $12 J/m2$. For such value of 
constant gg the energy of grain boundaries will diminish within the limits of $10$ $percents$ of its value 
for a relaxed boundary.

The terms in Eq. (\ref{b10}) proportional to $\bar{\lambda}_g$ and $\bar{\mu}_g$ must give at the same level of 
elastic stress of $180 MPa$, such contribution, when general effective energy of GB might not exceed the double 
energy of the free boundary (for copper $4 J/m^2$). It gives conditions for choosing constants in the limits
$\bar{\lambda}_g=0.25\cdot10^6 J/m^2$ and $\bar{\mu}_g=0.6\cdot10^6 J/m^2$. The values of these constants are 
different because the effects of shear are of greater significance for structural rearrangement of the solid.

Other constants are chosen issuing from the reasoning that the equilibrium (stationary) values of the density of 
grain boundaries were in the interval observed in experiments. For the grain boundaries these are two steady 
states with the density in a region $h_{g}^{st1} = 10 mm^{-1}$ and $h_{g}^{st2} = 10 \mu m^{-1}$, where the 
average grain size is $100 \mu m$ and $100 nm$, accordingly.

The same reasoning can be repeated for dislocations. The minimum excess energy of dislocations, in the absence of 
other factors for copper, equals approximately $\varphi_{0D}^*=5\cdot10^{-9} J/m$ \cite{k00}.

Using Eq. (\ref{b5}), which is true for both the equilibrium and non-equilibrium cases, one gets evolution 
Eqs (\ref{b6}) in the form
\begin{eqnarray}\label{b11}
\nonumber
\dfrac{\partial h_{D}}{\partial t}=-\gamma_{h_{D}}[\varphi_{1D}(h_{D}-h_{De})+\varphi_{gD}(h_{g}-h_{ge})],  \\
\dfrac{\partial h_{g}}{\partial t}=-\gamma_{h_{g}}[\varphi_{gD}(h_{D}-h_{De})+\varPhi(h_{g}-h_{ge})],
\end{eqnarray}
where
  \begin{equation}\label{b12}
\varphi=\varphi_{1g}-\varphi_{2g}(h_{g}+h_{ge})+\varphi_{3g}(h_{g}^2+h_{g}h_{ge}+h_{ge}^2).
  \end{equation}

As seen, the evolution equations do not directly depend on the parameters $\varphi_{0D}$ and $\varphi_{0g}$, 
but can depend on them through the equilibrium values $h_{De}$ and $h_{ge}$. For determination of this dependence 
it is necessary to find position of the maxima of PDF. As for basic defects, participating in SPD, this function 
is unknown; it creates some difficulties in application of the theory for calculation of specific systems. 
For overcoming the difficulties let us consider the method of effective potential in terms of the internal energy.

\subsection{Method of effective potential of the internal energy}

Let us suppose that the equilibrium energy of defect φle weakly depends on a current value of the density of defects, 
and it can be brought under the sign of differentiation in Eq. (\ref{b6}). Then one can introduce the effective 
internal energy
  \begin{equation}\label{b13}
\bar u =u - \sum_{l=1}^{N_{def}}\varphi_{le}h_{l}.
  \end{equation}

The evolution Eq. (\ref{b6}) assumes a form
  \begin{equation}\label{b14}
\dfrac{\partial h_{l}}{\partial t}=\pm\gamma_{l}\dfrac{\partial \bar u}{\partial h_{l}}.
  \end{equation}

Here the plus sign is selected in case if an equilibrium value $\varphi_{le}$ is in the region of convexity of 
the internal energy $u$, the minus sign is selected in the region of its concavity \cite{m10}. In the first case 
a stationary solution corresponds to a maximum of the effective energy, in the second case a stationary solution 
corresponds to its minimum. Formally, Eq. (\ref{b10}) realizes an extreme principle, as its stationary points 
coincide with a maximum or a minimum of the effective potential of the internal energy $\bar u$.

We take the effective energy   in the same form (\ref{b9}) as the initial internal energy u with the same 
coefficients of presentation (\ref{b10}) with the only difference that the equilibrium energy of $\varphi_{le}$ is 
included in coefficient of $\varphi_{0l}$, that is, $\varphi_{0l} > \varphi_{0l} - \varphi_{le}$ and $u > \bar u$.
Then, the set of evolution Eqs (\ref{b11}) can be written in the explicit form
\begin{eqnarray}\label{b15}
\dfrac{\partial h_{D}}{\partial t}=-\gamma_{h_{D}}(\varphi_{0D}-\varphi_{1D} h_{D}+\varphi_{gD} h_{g}),  \\
\nonumber
\dfrac{\partial h_{g}}{\partial t}=-\gamma_{h_{g}}(\varphi_{0g}-\varphi_{1g} h_{g}+\varphi_{2g} h_{g}^2
-\varphi_{3g} h_{g}^3+\varphi_{gD} h_{D}),
\end{eqnarray} 

Results got directly from the solution of the set of evolution Eqs (\ref{b11}), and by the method of effective 
potential of the internal energy (\ref{b15}), coincide, if relations between the coefficients of internal energy 
expansion $h_{ge}=(\varphi_{0g}-\varphi_{ge})/\varphi_{1g}$, $\varphi_{1g} >> \varphi_{2g}h_{ge}$ 
and $\varphi_{1g} >> \varphi_{3g}(h_{ge})^2$ are fulfilled.

Coming from the above analysis, such set of parameters and coefficients was accepted for calculations
$\lambda=\mu=2.08\cdot10^{10} Pa$, $\varphi_{0D}^*=5\cdot10^{-9} Jm^{-1}$, $\varphi_{1D}^*=10^{-24} Jm$, 
$g_{D}=2\cdot10^{-8} Jm^{-1}$, $\bar{\mu}_{D}=3.3\cdot10^{-4} Jm^{-1}$, $e_{D}=6\cdot10^{-23} Jm$, 
$\varphi_{0g}^*=0.4 Jm^{-2}$, $\varphi_{1g}^*=3\cdot10^{-6} Jm^{-1}$, $\varphi_{2g}=5.6\cdot10^{-13} J$, 
$\varphi_{3g}=3\cdot10^{-20} Jm$, $g_{g}=12 Jm^{-2}$, $\bar{\lambda}_{g}=2.5\cdot10^{5} Jm^{-2}$, 
$\bar{\mu}_{g}=6\cdot10^{5} Jm^{-2}$, $e_{g}=3.6\cdot10^{-4} Jm^{-1}$, $\varphi_{gD}=10^{-16} J$.

\begin{figure}
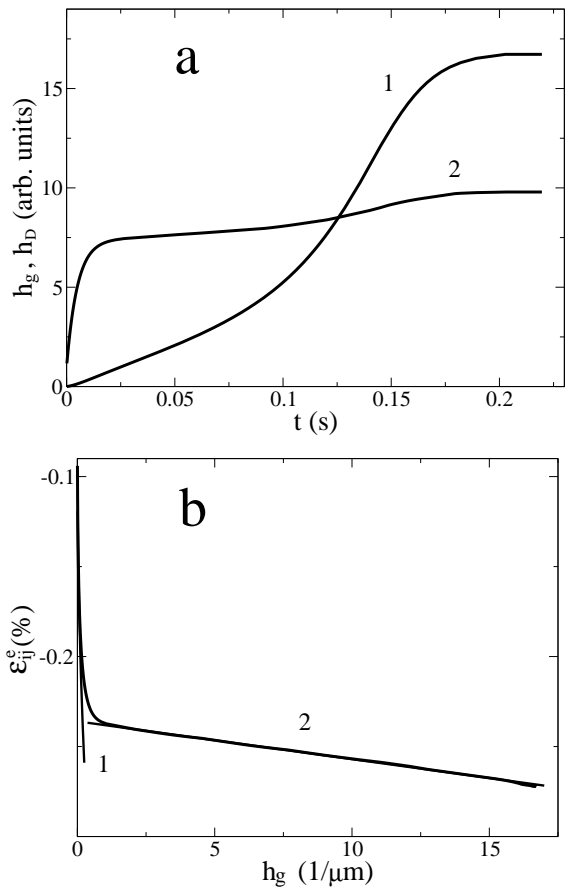

\hspace{0.06 cm}
\includegraphics [width=2.9 in] {fig_1a}
\vspace{0.26 cm}
\includegraphics [width=2.9 in] {fig_1b}
\caption{\label{f1} Regularities of defect-formation during SPD: a) kinetics of defects: 1 - density of grain boundaries; 
2 - density of dislocations; b) unified curve of strengthening: 1 - region of the Hall-Petch law; 
2 - region of the linear law of strengthening}
\end{figure}

Other coefficients in expressions for the internal energy are considered to be zero. A time step in numerical 
calculations is $\tau=0.67\cdot10^{-6}$, kinetic coefficients are $\gamma_{h_{D}} = 5\cdot10^{23} J m s$, 
$\gamma_{h_{g}} = 10^{6} J m^{-1} s$. The calculation of system evolution with these parameters and coefficients 
is shown in fig. \ref{f1}.

From fig. \ref{f1}а it is evident, that kinetics of grain boundaries and dislocations during the structural 
phase transition is closely correlated. At the first stage a growth in the number of dislocations initiates 
growth of grain boundaries and provokes the beginning of the structural phase transition. During the structural 
phase transition, when the density of dislocations has already gone on a stationary plateau, vice versa, the 
growth of grain boundaries provokes the growth of the density of dislocations. Thus, in this area, dislocations 
follow the grain boundaries in repeating the shape of curve of the structural phase transition, but to more weak extent.

\subsection{Strengthening curves}

As known the law of strengthening results from dislocation mobility decrease due to braking by different defects, 
as well as by dislocations from other slide planes. At the dislocation level the law of strengthening is described 
by Taylor relation \cite{m01,pf06}:
  \begin{equation}\label{b16}
\tau=\alpha \mu b\sqrt{h_{D}},
  \end{equation}
where $\tau$ is the shear stress; $\alpha$ is a coefficient, which takes on a value from the interval $0,2 \div 1,0$;
$\mu$ is the shear modulus; $b$ is the Burgers vector; $h_{D}$ is the density of dislocations (Here the modified 
denotation is used, which is general in a multilevel system of defects).

If GB are formed directly due to an outcoming of dislocations, in this case one can be limited by relations 
(\ref{b16}), and the Hall-Petch law at the grain level is got, as simple consequence from this relation acting 
at the dislocation level.

In the theory of NEET deformation is a control parameter which in the case of shear deformation relates to the stress 
by a simple dependence $\tau=\mu \varepsilon^e$ so, in terms of the theory the law of strengthening looks like
  \begin{equation}\label{b17}
\varepsilon^e=\alpha b\sqrt{h_{D}}.
  \end{equation}

This dependence can be used in NEET, as an additional relation to energetic and kinetic relations written above.

In the case of GB, a mode, which corresponds to a lower value of defectiveness, describes a coarse-grained structure, 
a mode, which corresponds to the greater value of defectiveness, describes a fine-grained structure. A possibility of 
forming different modes of the same defect is related to the microscopic mechanisms of deformation. In the case of 
grains they can be related to the circumstance that at the initial stage an increase of the general surface of GB is 
the effective mechanism of energy dissipation. The contribution of triple junctions can be ignored at this stage. 
The situation changes substantially, when the average size of grains decreases to $100$ $nm$. In this case, triple 
junctions can give a considerable contribution to the energy of boundaries, which can result in the formation of a 
new maximum in this region.

Triple junctions can be considered, as a specific type of defect, but as it is topologically attached to the 
grain boundaries, they can be integrated and considered, as one defect with a somewhat more difficult dependence 
of its energy on the number of defects.

The deformation stages during SPD are demonstrated by changes in the character of strengthening law at different 
stages of the process. At the first stage (area 1, fig. \ref{f1}b) the law of strengthening can be approximated 
by the Hall-Petch law, if expressed through elastic deformations
  \begin{equation}\label{b18}
\varepsilon^e=\varepsilon^{e0} - A \sqrt{h_{g}}.
  \end{equation}
where the constants are $\varepsilon^{e0}=-0.075$ and $A = 0.119mm$. The negative sign is taken because at 
compression that is in conditions typical of SPD the elastic deformation is negative. As seen from the picture, 
the elastic deformation changes within the limits of $0.1 \div 0.5$ $procents$, that in view of the value of 
shear modulus is in region of values for real materials.

In region 2, which corresponds to the most rapid phase of deformation during SPD, the law of strengthening 
can be approximated by a linear dependence
  \begin{equation}\label{b19}
\varepsilon^e=\varepsilon^{e1} - B h_{g}.
  \end{equation}
with constants $\varepsilon^{e1}=-0.27$ and $B=0.0059mm^2$. Just the same character of strengthening law 
change depending on deformation stage was noted in Ref. \cite{pf06}.

\section{Final remarks}

In the article, an alternative approach of non-equilibrium evolutional thermodynamics is considered and all 
features of the approach are demonstrated by the model example of solids with vacancies. We derive the system of 
kinetic equations in terms of the internal energy as the most fundamental thermodynamic potential. The generalized 
thermodynamic force immediately follows from differentiation of the probability distribution function during the 
procedure of finding the most probable state. In the equilibrium (or stationary) state such force is not zero, and 
associated with the equilibrium energy of defect in the equilibrium state.

The second feature which distinguishes this approach from traditional one is in using the density of defects 
as an independent thermodynamic variable instead of the configuration entropy. This allowed to modernize the writing 
of 1st law of thermodynamics, by adding to it an «entropy» terms in the form of product of the defect energy and 
the increment of the defect density (the last term in Eq. (\ref{b4})). In addition the first law of thermodynamics 
in this approach is written with dynamical transient phenomena taken into account in the form thermodynamic processes, 
to this end, the concepts of non-equilibrium temperature and non-equilibrium entropy were introduced (term before 
the last one in Eq. (\ref{b4})). Such a formulation can not be treated as fundamental, as the temperature and 
the entropy introduced in such a way will be determined by statistics of transient phenomena, which is not universal 
but strongly dependent on character of an external influence. At the same time, it can useful as an approximated 
relation for the solution of concrete problems.

Inclusion in the modified 1st law of thermodynamics of additional terms, describing internal non-equilibrium 
processes (they can be considered as analogues of the internal variables or the order parameters introduced 
by Landau), really, extends the dimensionality of the problem. The increase of problem dimensionality makes it 
simply certain. At the same time, we can introduce the generalized concept of the system state, depending on both 
equilibrium $\varepsilon_{ij}^e$, $S$ and non-equilibrium $\tilde{S}$ and $H_{l}$ variables.

Approach of NEET is applied to simulate the extraordinarily important for practical applications problem, 
the refining of grain structure of metals by severe plastic deformation methods. Within the framework of 
the two-level and two-mode approximation the system of kinetic equations, which includes the mutual kinetics 
of dislocations and grain boundaries, is deduced. For practical calculations the comfortable phenomenological 
model of the effective thermodynamic potential of the internal energy, which realizes the usual concept of 
the extreme state, is introduced.

Within the framework of this model the unified curve of strengthening, which has stage-like character, is 
constructed over the whole of the deformation interval. At the initial stage the curve of strengthening can be 
approximated by square-root dependence close to the Hall-Petch law, and at the finishing stage, it can be easily 
approximated by linear dependence.

\begin{acknowledgments}
This work is supported by the budget topic 0109U006004 of the NAS of Ukraine.
\end{acknowledgments}


\begin{thebibliography}{49}
\expandafter\ifx\csname natexlab\endcsname\relax\def\natexlab#1{#1}\fi
\expandafter\ifx\csname bibnamefont\endcsname\relax
  \def\bibnamefont#1{#1}\fi
\expandafter\ifx\csname bibfnamefont\endcsname\relax
  \def\bibfnamefont#1{#1}\fi
\expandafter\ifx\csname citenamefont\endcsname\relax
  \def\citenamefont#1{#1}\fi
\expandafter\ifx\csname url\endcsname\relax
  \def\url#1{\texttt{#1}}\fi
\expandafter\ifx\csname urlprefix\endcsname\relax\def\urlprefix{URL }\fi
\providecommand{\bibinfo}[2]{#2}
\providecommand{\eprint}[2][]{\url{#2}}


\bibitem[{\citenamefont{Herzfeld et~al.}(͑1928)\citenamefont{Herzfeld,
  and Rice}}]{hr28}
\bibinfo{author}{\bibfnamefont{K.~F.}~\bibnamefont{Herzfeld}}, \bibnamefont{and}
   \bibinfo{author}{\bibfnamefont{F.~O.}~\bibnamefont{Rice}},
  \bibinfo{journal}{Phys. Rev.} \textbf{\bibinfo{volume}{31}},
  \bibinfo{pages}{691} (\bibinfo{year}{͑1928}).

\bibitem[{\citenamefont{Rutgers et~al.}(͑1933)\citenamefont{Rutgers}}]{r33}
\bibinfo{author}{\bibfnamefont{A.~J.}~\bibnamefont{Rutgers}},
  \bibinfo{journal}{Ann. der Phys.} \textbf{\bibinfo{volume}{16}},
  \bibinfo{pages}{350} (\bibinfo{year}{͑1933}).

\bibitem[{\citenamefont{Kneser et~al.}(͑1933)\citenamefont{Kneser}}]{k33}
\bibinfo{author}{\bibfnamefont{H.~O.}~\bibnamefont{Kneser}},
  \bibinfo{journal}{J. Acoust. Soc. Am.} \textbf{\bibinfo{volume}{5}},\bibinfo{numer}{2},
  \bibinfo{pages}{122} (\bibinfo{year}{͑1933}).

\bibitem[{\citenamefont{Landau et~al.}(͑1936)\citenamefont{Landau,
  and Teller}}]{lt36}
\bibinfo{author}{\bibfnamefont{L.}~\bibnamefont{Landau}}, \bibnamefont{and}
   \bibinfo{author}{\bibfnamefont{E.}~\bibnamefont{Teller}},
  \bibinfo{journal}{Phys. Z. Sov.} \textbf{\bibinfo{volume}{10}},
  \bibinfo{pages}{34} (\bibinfo{year}{͑1936}).

\bibitem[{\citenamefont{Mandelstam et~al.}(͑1937)\citenamefont{Mandelstam,
  and Teller}}]{ml37}
\bibinfo{author}{\bibfnamefont{L.}~\bibnamefont{Mandelstam}}, \bibnamefont{and}
   \bibinfo{author}{\bibfnamefont{M.}~\bibnamefont{Leontovich}},
  \bibinfo{journal}{J. Theor. Experim. Phys.} \textbf{\bibinfo{volume}{7}},
  \bibinfo{pages}{438} (\bibinfo{year}{͑1937}).

\bibitem[{\citenamefont{Landau et~al.}(͑1937)\citenamefont{Landau}}]{l37}
\bibinfo{author}{\bibfnamefont{L.}~\bibnamefont{Landau}},
  \bibinfo{journal}{J. Theor. Experim. Phys.} \textbf{\bibinfo{volume}{7}},
  \bibinfo{pages}{19} (\bibinfo{year}{͑1937}).

\bibitem[{\citenamefont{Landau et~al.}(͑1954)\citenamefont{Landau,
  and Khalatnikov}}]{lk54}
\bibinfo{author}{\bibfnamefont{L.}~\bibnamefont{Landau}}, \bibnamefont{and}
   \bibinfo{author}{\bibfnamefont{I.}~\bibnamefont{Khalatnikov}},
  \bibinfo{journal}{Sov. Physics Doklady} \textbf{\bibinfo{volume}{96}},
  \bibinfo{pages}{459} (\bibinfo{year}{͑1954}).

\bibitem[{\citenamefont{Ginzburg et~al.}(͑1950)\citenamefont{Ginzburg,
  and Landau}}]{gl50}
\bibinfo{author}{\bibfnamefont{L.}~\bibnamefont{Ginzburg}}, \bibnamefont{and}
   \bibinfo{author}{\bibfnamefont{L.}~\bibnamefont{Landau}},
  \bibinfo{journal}{J. Theor. Experim. Phys.} \textbf{\bibinfo{volume}{20}},
  \bibinfo{pages}{1064} (\bibinfo{year}{͑1950}).

\bibitem[{\citenamefont{Aranson et~al.}(2000)\citenamefont{Aranson, Kalatsky,
  and Vinokur}}]{akv00}
\bibinfo{author}{\bibfnamefont{I.~S.}~\bibnamefont{Aranson}},
  \bibinfo{author}{\bibfnamefont{V.~A.}~\bibnamefont{Kalatsky}}, \bibnamefont{and}
  \bibinfo{author}{\bibfnamefont{V.~M.}~\bibnamefont{Vinokur}},
  \bibinfo{journal}{Phys. Rev. Let.} \textbf{\bibinfo{volume}{85}},
  \bibinfo{pages}{118} (\bibinfo{year}{2000}).

\bibitem[{\citenamefont{Karma et~al.}(2001)\citenamefont{Karma, Kessler,
  and Levine}}]{kkl01}
\bibinfo{author}{\bibfnamefont{A.}~\bibnamefont{Karma}},
  \bibinfo{author}{\bibfnamefont{D.~A.}~\bibnamefont{Kessler}}, \bibnamefont{and}
  \bibinfo{author}{\bibfnamefont{H.} \bibnamefont{Levine}},
  \bibinfo{journal}{Phys. Rev. Let.} \textbf{\bibinfo{volume}{87}},
  \bibinfo{pages}{045501} (\bibinfo{year}{2001}).

\bibitem[{\citenamefont{Eastgate et~al.}(2002)\citenamefont{Eastgate, Sethna,
  Rauscher, and Cretegny}}]{esrc02}
\bibinfo{author}{\bibfnamefont{L.~O.}~\bibnamefont{Eastgate}},
  \bibinfo{author}{\bibfnamefont{J.~P.}~\bibnamefont{Sethna}},
  \bibinfo{author}{\bibfnamefont{M.}~\bibnamefont{Rauscher}},
  \bibinfo{author}{\bibfnamefont{T.}~\bibnamefont{Cretegny}},
  \bibinfo{author}{\bibfnamefont{C.~S.}~\bibnamefont{Chen}}, \bibnamefont{and}
  \bibinfo{author}{\bibfnamefont{C.~R.}~\bibnamefont{Myers}},
  \bibinfo{journal}{Phys. Rev. E} \textbf{\bibinfo{volume}{65}},
  \bibinfo{pages}{036117} (\bibinfo{year}{2002}).

\bibitem[{\citenamefont{Levitas et~al.}(2002)\citenamefont{Levitas, Preston,
  and Lee}}]{lpl02}
\bibinfo{author}{\bibfnamefont{V.~I.}~\bibnamefont{Levitas}},
  \bibinfo{author}{\bibfnamefont{D.~L.}~\bibnamefont{Preston}}, \bibnamefont{and}
  \bibinfo{author}{\bibfnamefont{D.~W.} \bibnamefont{Lee}},
  \bibinfo{journal}{Phys. Rev. B} \textbf{\bibinfo{volume}{68}},
  \bibinfo{pages}{134201} (\bibinfo{year}{2003}).

\bibitem[{\citenamefont{Ramirez et~al.}(2004)\citenamefont{Ramirez, Beckermann,
  Karma, and Diepers}}]{rbkd04}
\bibinfo{author}{\bibfnamefont{J.~C.}~\bibnamefont{Ramirez}},
  \bibinfo{author}{\bibfnamefont{C.}~\bibnamefont{Beckermann}},
  \bibinfo{author}{\bibfnamefont{A.}~\bibnamefont{Karma}}, \bibnamefont{and}
  \bibinfo{author}{\bibfnamefont{H.~J.} \bibnamefont{Diepers}},
  \bibinfo{journal}{Phys. Rev. E} \textbf{\bibinfo{volume}{69}},
  \bibinfo{pages}{051607} (\bibinfo{year}{2004}).

\bibitem[{\citenamefont{Achim et~al.}(2006)\citenamefont{Achim, Karttunen,
  Elder, Granato, Ala-Nissila, and Ying}}]{akegny06}
\bibinfo{author}{\bibfnamefont{C.~V.}~\bibnamefont{Achim}},
  \bibinfo{author}{\bibfnamefont{M.}~\bibnamefont{Karttunen}},
  \bibinfo{author}{\bibfnamefont{K.~R.}~\bibnamefont{Elder}},
  \bibinfo{author}{\bibfnamefont{E.}~\bibnamefont{Granato}},
  \bibinfo{author}{\bibfnamefont{T.}~\bibnamefont{Ala-Nissila}},
  \bibnamefont{and} \bibinfo{author}{\bibfnamefont{S.~C.} \bibnamefont{Ying}},
  \bibinfo{journal}{Phys. Rev. E} \textbf{\bibinfo{volume}{74}},
  \bibinfo{pages}{021104} (\bibinfo{year}{2006}).

\bibitem[{\citenamefont{Svandal}(2006)}]{s06}
\bibinfo{author}{\bibfnamefont{A.} \bibnamefont{Svandal}},
  \emph{\bibinfo{title}{Modeling hydrate phase transitions using mean-field approaches}}
  (\bibinfo{publisher}{University of Bergen}, \bibinfo{address}{Bergen},
  \bibinfo{year}{2006}).

\bibitem[{\citenamefont{Rosam et~al.}(2009)\citenamefont{Rosam, Jimack, and
  Mullis}}]{rjm09}
\bibinfo{author}{\bibfnamefont{J.}~\bibnamefont{Rosam}},
  \bibinfo{author}{\bibfnamefont{P.~K.}~\bibnamefont{Jimack}}, \bibnamefont{and}
  \bibinfo{author}{\bibfnamefont{A.~M.}~\bibnamefont{Mullis}},
  \bibinfo{journal}{Phys. Rev. E} \textbf{\bibinfo{volume}{79}},
  \bibinfo{pages}{030601(R)} (\bibinfo{year}{2009}).

\bibitem[{\citenamefont{Coleman et~al.}(͑1967)\citenamefont{Coleman,
  and Gurtin}}]{cg67}
\bibinfo{author}{\bibfnamefont{B.~D.}~\bibnamefont{Coleman}}, \bibnamefont{and}
   \bibinfo{author}{\bibfnamefont{M.~E.}~\bibnamefont{Gurtin}},
  \bibinfo{journal}{J. Chem. Phys.} \textbf{\bibinfo{volume}{47}},
  \bibinfo{pages}{597} (\bibinfo{year}{͑1967}).

\bibitem[{\citenamefont{Dyarmati}(1970)}]{d70}
\bibinfo{author}{\bibfnamefont{I.} \bibnamefont{Dyarmati}},
  \emph{\bibinfo{title}{Non-equilibrium thermodynamics. Field. theory and variational principles}} 
(\bibinfo{publisher}{Springer}, \bibinfo{address}{Heidelberg}, 
\bibinfo{year}{1970}).

\bibitem[{\citenamefont{Petrov and Brankov}(1986)}]{pb69}
\bibinfo{author}{\bibfnamefont{N.~P.}~\bibnamefont{Petrov}} \bibnamefont{and}
  \bibinfo{author}{\bibfnamefont{J.~G.}~\bibnamefont{Brankov}},
  \emph{\bibinfo{title}{Modern problems of thermodynamics}} (\bibinfo{publisher}{«Mir»}, 
\bibinfo{address}{Moscow}, \bibinfo{year}{1986}).

\bibitem[{\citenamefont{Metlov et~al.}(͑1997)\citenamefont{Metlov,
  and Morozov}}]{m97}
\bibinfo{author}{\bibfnamefont{L.~S.}~\bibnamefont{Metlov}}, \bibnamefont{and}
   \bibinfo{author}{\bibfnamefont{A.~F.}~\bibnamefont{Morozov}},
  \bibinfo{journal}{High Press. Phys. Technics} \textbf{\bibinfo{volume}{7}}, \bibinfo{numer}{3},
  \bibinfo{pages}{58} (\bibinfo{year}{͑1997}).


\bibitem[{\citenamefont{Van et~al.}(2001)\citenamefont{Van}}]{v01}
\bibinfo{author}{\bibfnamefont{P.}~\bibnamefont{Van}},
  \bibinfo{journal}{J. Non-Equilib. Thermodyn.} \textbf{\bibinfo{volume}{26}},  \bibinfo{numer}{2},
  \bibinfo{pages}{167} (\bibinfo{year}{2001}).

\bibitem[{\citenamefont{Van et~al.}(2002)\citenamefont{Van}}]{v02}
\bibinfo{author}{\bibfnamefont{P.}~\bibnamefont{Van}},
  \bibinfo{journal}{Technische Mechanic} \textbf{\bibinfo{volume}{22}},  \bibinfo{numer}{2},
  \bibinfo{pages}{104} (\bibinfo{year}{2002}).

\bibitem[{\citenamefont{Kluev et~al.}(2005)\citenamefont{Kluev,
  and Trusov}}]{kt05}
\bibinfo{author}{\bibfnamefont{A.~V.}~\bibnamefont{Kluev}}, \bibnamefont{and}
   \bibinfo{author}{\bibfnamefont{P.~V.}~\bibnamefont{Trusov}},
  \bibinfo{journal}{Mathematical simulation of systems and processes} \textbf{\bibinfo{volume}{}}\bibinfo{numer}{13},
  \bibinfo{pages}{29} (\bibinfo{year}{2005}).


\bibitem[{\citenamefont{Knjazeva et~al.}(2005)\citenamefont{Knjazeva}}]{k05}
\bibinfo{author}{\bibfnamefont{A.~G.}~\bibnamefont{Knjazeva}},
  \bibinfo{journal}{Mathematical simulation of systems and processes} \textbf{\bibinfo{volume}{}}\bibinfo{numer}{13},
  \bibinfo{pages}{45} (\bibinfo{year}{2005}).

\bibitem[{\citenamefont{Reguera et~al.}(1998)\citenamefont{Reguera, Rubi,
  and Perez-Madrid}}]{rrp98}
\bibinfo{author}{\bibfnamefont{D.}~\bibnamefont{Reguera}},
  \bibinfo{author}{\bibfnamefont{J.~M.}~\bibnamefont{Rubi}}, \bibnamefont{and}
  \bibinfo{author}{\bibfnamefont{A.}~\bibnamefont{Perez-Madrid}},
  \bibinfo{journal}{Physica A} \textbf{\bibinfo{volume}{259}},
  \bibinfo{pages}{10} (\bibinfo{year}{1998}).

\bibitem[{\citenamefont{Rubi et~al.}(2003)\citenamefont{Rubi,
  and Gadomski}}]{rg03}
\bibinfo{author}{\bibfnamefont{J.~M.}~\bibnamefont{Rubi}}, \bibnamefont{and}
   \bibinfo{author}{\bibfnamefont{A.}~\bibnamefont{Gadomski}},
  \bibinfo{journal}{Physica A} \textbf{\bibinfo{volume}{326}},
  \bibinfo{pages}{333} (\bibinfo{year}{2003}).

\bibitem[{\citenamefont{Reguera et~al.}(2005)\citenamefont{Reguera, Vilar,
  and Rubi}}]{rvr05}
\bibinfo{author}{\bibfnamefont{D.}~\bibnamefont{Reguera}},
  \bibinfo{author}{\bibfnamefont{J.~M.~G.}~\bibnamefont{Vilar}}, \bibnamefont{and}
  \bibinfo{author}{\bibfnamefont{J.~M.}~\bibnamefont{Rubi}},
  \bibinfo{journal}{J. Phys. Chem.B} \textbf{\bibinfo{volume}{109}},
  \bibinfo{pages}{21502} (\bibinfo{year}{2005}).

\bibitem[{\citenamefont{Pérez-Madrid, Rubi and Lapas}(2011)}]{pmrl11}
\bibinfo{author}{\bibfnamefont{A.}~\bibnamefont{Pérez-Madrid}},
\bibinfo{author}{\bibfnamefont{J.~M.}~\bibnamefont{Rubi}}, \bibnamefont{and}
  \bibinfo{author}{\bibfnamefont{L.~C.}~\bibnamefont{Lapas}},
  \emph{\bibinfo{title}{Mesoscopic non-equilibrium thermodynamics: application to radiative heat exchange 
in nanostructures, in book: Thermodynamics M. Tadasi (Ed) p. 195-204}} 
(\bibinfo{publisher}{InTech}, 
\bibinfo{address}{Germany}, \bibinfo{year}{2011}).

\bibitem[{\citenamefont{Metlov}(2007)}]{m07}
\bibinfo{author}{\bibfnamefont{L.~S.}~\bibnamefont{Metlov}}, 
  \bibinfo{journal}{Bulletin of Donetsk Univ. Ser. A: Nature Sciences}, \bibinfo{numer}{1},
  \bibinfo{pages}{167} (\bibinfo{year}{2007}).

\bibitem[{\citenamefont{Metlov}(2007)}]{m07c}
\bibinfo{author}{\bibfnamefont{L.~S.}~\bibnamefont{Metlov}},
  \emph{\bibinfo{title}{Proceedings of the 2nd international conference «Deformation and Fracture of Materials 
and Nanomaterials» DFMN2007, p. 642-644}}
(\bibinfo{publisher}{IMET RAN}, 
\bibinfo{address}{Moscow}, \bibinfo{year}{2007}).

\bibitem[{\citenamefont{Steinberg}(1989)}]{s89}
\bibinfo{author}{\bibfnamefont{A.~S}~\bibnamefont{Steinberg}},
  \emph{\bibinfo{title}{Reportage from alloys world}} (\bibinfo{publisher}{PhysMatLit},
  \bibinfo{address}{Moscow}, \bibinfo{year}{1989}).

\bibitem[{\citenamefont{Landau and Lifshitz}(1969)}]{ll69}
\bibinfo{author}{\bibfnamefont{L.}~\bibnamefont{Landau}} \bibnamefont{and}
  \bibinfo{author}{\bibfnamefont{E.}~\bibnamefont{Lifshitz}},
  \emph{\bibinfo{title}{Statistical mechanics}} (\bibinfo{publisher}{Pergamon
  Press}, \bibinfo{address}{Oxford}, \bibinfo{year}{1969}).

\bibitem[{\citenamefont{Metlov}(2010)}]{m10}
\bibinfo{author}{\bibfnamefont{L.~S}~\bibnamefont{Metlov}},
  \bibinfo{journal}{Phys. Rev. E} \textbf{\bibinfo{volume}{81}},
  \bibinfo{pages}{051121} (\bibinfo{year}{2010}).

\bibitem[{\citenamefont{Metlov}()}]{m10p}
\bibinfo{author}{\bibfnamefont{L.~S.} \bibnamefont{Metlov}},
  \eprint{cond-mat/1003.0450}.

\bibitem[{\citenamefont{Metlov}(2011)}]{m11}
\bibinfo{author}{\bibfnamefont{L.~S}~\bibnamefont{Metlov}},
  \bibinfo{journal}{Phys. Rev. Lett.} \textbf{\bibinfo{volume}{106}},
  \bibinfo{pages}{165506} (\bibinfo{year}{2011}).

\bibitem[{\citenamefont{Gufan}(1997)}]{g97}
\bibinfo{author}{\bibfnamefont{Yu.~M.}~\bibnamefont{Gufan}},
  \bibinfo{journal}{Soros education journal} \textbf{\bibinfo{numer}{7}},
  \bibinfo{pages}{109} (\bibinfo{year}{1997}).

\bibitem[{\citenamefont{Bazarov}(1986)}]{b64}
\bibinfo{author}{\bibfnamefont{N.~P.}~\bibnamefont{Bazarov}},
  \emph{\bibinfo{title}{Thermodynamics}} (\bibinfo{publisher}{Pergamon}, 
\bibinfo{address}{Oxford}, \bibinfo{year}{1964}).

\bibitem[{\citenamefont{Metlov}(2007)}]{m09}
\bibinfo{author}{\bibfnamefont{L.~S.}~\bibnamefont{Metlov}}, 
  \bibinfo{journal}{Bulletin of Donetsk Univ. Ser. A: Nature Sciences}, \bibinfo{numer}{2},
  \bibinfo{pages}{144} (\bibinfo{year}{2009}).

\bibitem[{\citenamefont{Kochegarov}(2000)\citenamefont{Kochegarov}}]{k00}
\bibinfo{author}{\bibfnamefont{G.~G.}~\bibnamefont{Kochegarov}},
  \bibinfo{journal}{Technical Physics Letters} \textbf{\bibinfo{volume}{26}}, \bibinfo{numer}{11},
  \bibinfo{pages}{41} (\bibinfo{year}{2000}).

\bibitem[{\citenamefont{Podrezov}(2006)\citenamefont{Podrezov}}]{pf06}
\bibinfo{author}{\bibfnamefont{Yu.~N.}~\bibnamefont{Podrezov}},\bibnamefont{and}
\bibinfo{author}{\bibfnamefont{S.~A.}~\bibnamefont{Firstov}},
  \bibinfo{journal}{High Press. Phys. Technics} \textbf{\bibinfo{volume}{16}}, \bibinfo{numer}{4},
  \bibinfo{pages}{58} (\bibinfo{year}{2006}).

\bibitem[{\citenamefont{Moiseev}(2001)\citenamefont{Moiseev}}]{m01}
\bibinfo{author}{\bibfnamefont{V.~F.}~\bibnamefont{Moiseev}},
  \bibinfo{journal}{Metal Physics and Advanced Technologies} \textbf{\bibinfo{volume}{23}}, \bibinfo{numer}{3},
  \bibinfo{pages}{387} (\bibinfo{year}{2001}).






\end{thebibliography}
\end{document}